\documentstyle[12pt]{article}% 11pt, article report letter
\begin{document}
\renewcommand{\thefootnote}{\fnsymbol{footnote}}
\begin{flushright}
TIFR/TH/97-61\\
\end{flushright}
\vspace*{3cm}
\begin{center}
{\Large \bf \boldmath Aspects of hadronic $J/\psi$ production
\footnote{Presented at the Pacific Particle Physics Phenomenology ($P^4$ '97)
Conference at the Seoul National University, S.~Korea, October 31 - 
November 2, 1997.}} \\
\vskip 32pt
{\bf K. Sridhar\footnote{sridhar@theory.tifr.res.in}}\\
{\it Theory Group, Tata Institute of Fundamental Research, \\ 
Homi Bhabha Road, Bombay 400 005, India.}

\vspace{100pt}
{\bf ABSTRACT}
\end{center}
\vspace{12pt}

In this talk, I start with a brief introduction to 
Non-Relativistic QCD (NRQCD) and its applications to
quarkonium physics. This theory has provided a consistent framework
for the physics of quarkonia, in particular, the colour-octet Fock
components predicted by NRQCD have important implications for
the phenomenology of charmonium production in experiments. 
We discuss the applications of NRQCD to $J/\psi$ production
at Tevatron and the tests of the theory in other experiments.
Finally we discuss the production of the ${}^1P_1$ charmonium state, $h_c$, 
at large-$p_T$ at the Tevatron. The observation of $h_c$ is interesting, 
since its
existence has yet to be experimentally confirmed. Moreover, this
rate is a $prediction$ of NRQCD, and the observation of
the $h_c$ at the Tevatron can, therefore, be used as a test 
of NRQCD.

\vfill
\clearpage
\pagestyle{empty}
$J/\psi$ production has traditionally been described in
terms of the colour-singlet model \cite{berjon, br}, which, 
in spite of its reasonable success with its predictions at lower
energies, fails completely in explaining the data on 
large-$p_T$ $J/\psi$ production at the Tevatron \cite{cdf}. 
This has led to a major revision of the theory of quarkonium formation. 
In the colour-singlet model, the relative velocity, $v$ 
between the charm quarks is neglected. However, for charmonia
$v^2 \sim 0.3$, and is not negligible. It is possible to include the 
effects coming from higher orders in a perturbation series in $v$ 
in non-relativistic QCD (NRQCD) \cite{caswell, bbl}. 

In this picture, the physical charmonium state can be expanded 
in terms of its Fock-components in a perturbation series in $v$, 
and the $c \bar c$ states appear in either colour-singlet or 
colour-octet configurations in this series. The 
colour-octet $c \bar c$ state makes a transition into a physical
state by the emission of one or more soft gluons, 
dominantly $via$ $E1$ or $M1$ transitions with
higher multipoles being suppressed by powers of $v$. 
The cross-section for the production of a meson $H$ 
then takes on the following factorised form:
\begin{equation}
   \sigma(H)\;=\;\sum_{n=\{\alpha,S,L,J\}} {F_n\over m^{d_n-4}}
       \langle{\cal O}^H_\alpha({}^{2S+1}L_J)\rangle
\label{e1}
\end{equation}
where $F_n$'s are the short-distance coefficients and ${\cal O}_n$ are local
4-fermion operators, of naive dimension $d_n$, describing the long-distance
physics. 

In fact, the importance of the colour-octet components was first noted
\cite{bbl2} in the case of $P$-wave charmonium decays, and even in
the production case the importance of these components was first
seen \cite{jpsi} in the production of $P$-state charmonia at the
Tevatron. The surprise was that even for the production of $S$-states
such as the $J/\psi$ or $\psi'$, where the colour singlet components give
the leading contribution in $v$, the inclusion of sub-leading octet states
was seen to be necessary for phenomenological reasons \cite{brfl}. 
While the inclusion of the colour-octet components seem to be
necessiated by the Tevatron charmonium data, the normalisation of
these data cannot be predicted because the the long-distance matrix 
elements are not calculable. The data allow a linear combination of 
octet matrix-elements to be fixed \cite{cgmp,cho}, and much effort
has made recently to understand the implications of these colour-octet
channels for $J/\psi$ production in other experiments. We discuss
some of these below.

\begin{enumerate} 
\item
The prediction \cite{lep1, lep2, brcyu} for prompt $J/\psi$ production 
at LEP in the colour-singlet model is of the 
order of $3 \times 10^{-5}$, which is almost an order of magnitude 
below the experimental number for the branching fraction obtained from 
LEP \cite{lep3}. Recently, the colour-octet contributions
to $J/\psi$ production in this channel have been studied \cite{lep5,lep6}
and it is found that the inclusion of the colour-octet contributions 
in the fragmentation functions results in a predictions for the 
branching ratio which is $1.4 \times 10^{-4}$ which is compatible
with the measured values of the branching fraction from LEP \cite{lep3}.

\item
The production of $J/\psi$ in low energy $e^+ e^-$ machines can
also provide a stringent test of the colour-octet mechanism \cite{brch}.
In this case, the colour-octet contributions dominate near the upper
endpoint of the $J/\psi$ energy spectrum, and the signature for
the colour-octet process is a dramatic change in the angular distribution
of the $J/\psi$ near the endpoint. 

\item
One striking prediction of the colour-octet fragmentation process both
for $p \bar p$ colliders and for $J/\psi$ production at the $Z$-peak,
is that the $J/\psi$ coming from the process $g \rightarrow J/\psi X$
is produced in a transversely polarised state \cite{trans}. For the 
colour-octet $c \bar c$ production, this is predicted to be a 100\% 
transverse polarisation, and heavy-quark spin symmetry will then ensure 
that non-perturbative effects which convert the $c \bar c$ to a $J/\psi$
will change this polarisation only very mildly. This spin-alignment
can, therefore, be used as a test of colour-octet fragmentation.

\item
The colour-octet components are found \cite{hadro} to dominate the 
production processes in fixed-target $pp$ and $\pi p$ experiments.
Using the colour-octet matrix elements extracted from elastic photoproduction 
data it is possible to get a very good description of the 
$\sqrt{s}$-dependence and also the $x_F$ and rapidity distributions.

\item
The associated production of a $J/\psi + \gamma$ is also a 
crucial test of the colour-octet components \cite{kim} and
also of the fragmentation picture \cite{psigam}. Similar
tests can be concieved of with double $J/\psi$ production 
at the Tevatron \cite{bfp}. 

\item
$J/\psi$ and $\psi^{\prime}$ production in $pp$ collisions at 
centre-of-mass energies of 14~TeV at the LHC also provides a 
crucial test of colour-octet fragmentation \cite{lhc}. 

\item
The fragmentation process has also definite predictions for the
large-$p_T$ inelastic $J/\psi$ photoproduction at HERA \cite{hera}. 
At HERA, the colour-singlet component dominates inelastic $J/\psi$
production but at large-$z$ ($z>0.7$) the colour-octet components
start growing rapidly. Such a rapid growth is clearly in conflict 
with experiment \cite{cakr}. The applicability of NRQCD factorisation
at these large values of $z$ has been questioned \cite{brw}, and
more work needs to be done before the disagreement between theory
and the HERA inelastic data can be interpreted as a failure of
NRQCD.

\end{enumerate} 

One important feature of the NRQCD Lagrangian is that it shows
an approximate heavy-quark symmetry, which is valid to $O(v^2) \sim 0.3$.
The implication of this symmetry is that the nonperturbative parameters
have a weak dependence on the magnetic quantum number. We
show that the production cross-section of the ${}^1P_1$ charmonium 
state, $h_c$, can be $predicted$ in NRQCD, because of the heavy-quark
symmetry. The Tevatron data on $\chi_c$ production fixes \cite{cho} the 
colour-octet matrix element which specifies the transition of a 
${}^3S_1$ octet state into a ${}^3P_J$ state. We would expect from 
heavy-quark spin symmetry of the NRQCD Lagrangian that the 
matrix-element for ${}^1S_0^{\lbrack 8 \rbrack} \rightarrow h_c$ should be
of the same order as that for ${}^3S_1^{\lbrack 8 \rbrack} \rightarrow 
{}^3P_1$. The production of the ${}^1P_1$ charmonium state, $h_c$,
is, therefore, a crucial prediction of NRQCD.

The production of the $h_c$ is interesting in its own right~: 
charmonium spectroscopy \cite{onep} predicts this state to exist at 
the centre-of-gravity of the $\chi_c ({}^3P_J)$ states. While the E760 
collaboration at the Fermilab has reported \cite{e760} the first 
observation of this resonance its existence needs further confirmation. 
In the following, we report on the calculations for $h_c$ production 
at the Tevatron presented in Ref.~\cite{mine}.

Being a $P$-state, the leading colour-singlet contribution is
already at $O(v^2)$, and at the same order we have the octet
production of the ${}^1P_1$ state through an intermediate 
${}^1S_0$ state. The octet channel is, therefore, expected to
be very important for the production of this state. Once produced, the
$h_c$ can be detected by its decay into a $J/\psi + \pi^0$.
The subprocesses that we are interested in are the following:
\begin{eqnarray}
g + g &\rightarrow {}^1P_1^{\lbrack 1 \rbrack} + g, \nonumber \\
g + g &\rightarrow {}^1S_0^{\lbrack 8 \rbrack} + g, \nonumber \\
q(\bar q)+ g &\rightarrow {}^1S_0^{\lbrack 8 \rbrack } + 
                  q(\bar q), \nonumber \\
q + \bar q &\rightarrow {}^1S_0^{\lbrack 8 \rbrack } + g.
      \label{e2}
\end{eqnarray}
The ${}^1S_0^{\lbrack 8 \rbrack} \rightarrow h_c$ is mediated by
a gluon emission in a $E1$ transition. 

The cross-section $d\sigma / dp_T$ for $h_c$ production at the 
Tevatron energy ($\sqrt{s}= 1.8$~TeV) presented in Ref.~\cite{mine} 
are shown in Fig.~1. For 20~pb${}^{-1}$ total luminosity, for $p_T$
integrated between 5 and 20~GeV we expect of the order of 650
events in the $J/\psi + \pi$ channel. Of these, the contribution 
from the colour-singlet channel is a little more than 40, while the 
octet channel gives more than 600 events. The colour-octet dominance 
is more pronounced at large-$p_T$. Recent results on $J/\psi$ production
from CDF are based on a total luminosity of 110~pb${}^{-1}$. For
this sample, more than 3000 events can be expected to come
from the decay of the $h_c$ into a $J/\psi$ and a $\pi$. With this
large event rate, the $h_c$ should certainly be observable if the
$pi^0$ coming from its decay can be reconstructed efficiently. 

In conclusion, we find that a reasonably large rate for the production
of the ${}^1P_1$ is expected at the Tevatron, with a dominant contribution
from the colour-octet production channel. Heavy-quark symmetry relations
allow us to infer the size of the non-perturbative matrix-elements and 
to predict the rate for $h_c$ production and its subsequent
decay into a $J/\psi$ and a $\pi$. By looking at $J/\psi$ events associated
with a soft pion, it should be possible to pin down the elusive ${}^1P_1$
resonance at the Tevatron, given the large number of $J/\psi$ events
already available. We believe that this is a firm prediction of the
NRQCD framework and may be a useful way of distinguishing this from
other models of quarkonium formation.

\begin{figure}[h]
\vskip 6.2in\relax\noindent
          \relax{\includegraphics{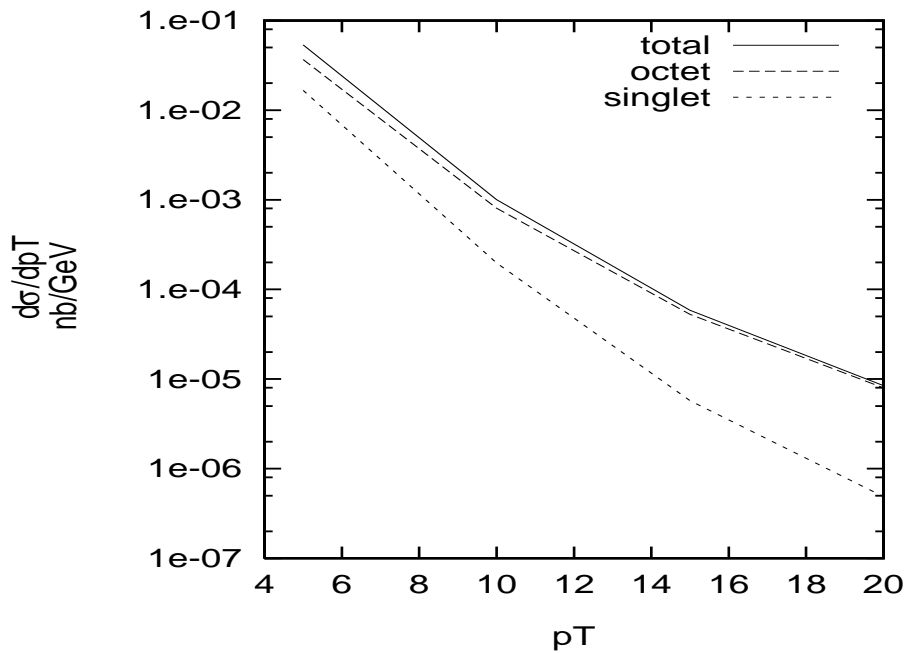}}  
\vspace{-40ex}
\caption{$d\sigma/dp_T$ (in nb/GeV) for $h_c$ production (after folding in
with Br($h_c \rightarrow J/\psi+\pi$)=0.5) at
1.8~TeV c.m. energy with $-0.6 \le y \le 0.6$. }
\end{figure}

%\clearpage


\begin{thebibliography}{999}

\bibitem{berjon} E.L.~Berger and D.~Jones, {\it Phys. Rev.} {\bf D 23}
(1981) 1521.

\bibitem{br} R.~Baier and R.~R\" uckl, {\it Z. Phys.} {\bf C 19}
(1983) 251.

\bibitem{cdf} F.~Abe et al., {\it Phys. Rev. Lett.} {\bf 69}
(1992) 3704; {\it Phys. Rev. Lett.} {\bf 71} (1993) 2537; 
R.~Demina, FERMILAB-CONF-96-201-E, Presented at the 11th Topical
Workshop on Proton-Antiproton Collider Physics (PBARP 96), Padua,
Italy, 26th May -- 1st June 1996.

\bibitem{caswell} W.E.\ Caswell and G.P.\ Lepage, 
  {\it Phys. Lett.}, {\bf B 167} (1986) 437.

\bibitem{bbl} G.T.~Bodwin, E.~Braaten and G.P.~Lepage, {\it Phys. Rev.} 
{\bf D 51} (1995) 1125.

\bibitem{bbl2} G.T.~Bodwin, E.~Braaten and G.P.~Lepage, {\it Phys. Rev.} 
{\bf D 46} (1992) R1914.

\bibitem{jpsi} E.~Braaten, M.A.~Doncheski, S.~Fleming and M.~Mangano,
{\it Phys. Lett.} {\bf B 333} (1994) 548; D.P.~Roy and K.~Sridhar, 
{\it Phys. Lett.} {\bf B 339} (1994) 141; M.~Cacciari and M.~Greco, 
{\it Phys. Rev. Lett.} {\bf 73} (1994) 1586. 

\bibitem{brfl} E.~Braaten and S.~Fleming, {\it Phys. Rev. Lett.}
{\bf 74} (1995) 3327.

\bibitem{cgmp} M.~Cacciari, M.~Greco, M.~Mangano and A.~Petrelli,
{\it Phys. Lett.} {\bf B 356} (1995) 553.  

\bibitem{cho}
   P. Cho and A.K.~Leibovich, {\it Phys. Rev.}, {\bf D 53} (1996) 150;
   {\it Phys. Rev.}, {\bf D 53} (1996) 6203.

\bibitem{lep1} V.~Barger, K.~Cheung and W.-Y.~Keung, 
{\it Phys. Rev.} {\bf D 41} (1990) 1541.

\bibitem{lep2} W.-Y.~Keung, {\it Phys. Rev.} {\bf D 23} (1981) 2072;
J.H.~K\" uhn and H.~Schneider, {\it Phys. Rev.} {\bf D 24} (1981) 2996.

\bibitem{brcyu} E.~Braaten, K.Cheung and T.C.~Yuan,
{\it Phys. Rev.} {\bf D 48} (1993) 4230.

\bibitem{lep3} O.~Adriani et al. (L3 collaboration), {\it Phys. Lett.} 
{\bf B 288} (1992) 412; 
P.~Abreu et al. (DELPHI collaboration), {\it Phys. Lett.} 
{\bf B 341} (1994) 109; 
G.~Alexander et al. (OPAL collaboration), CERN Preprint CERN-PPE-95-153.

\bibitem{lep5} K.~Cheung, W.-Y.~Keung and T.C.~Yuan, 
{\it Phys. Rev. Lett.} {\bf 76} (1996) 877;
S.~Baek, P.~Ko, J.~Lee and H.S.~Song, 
{\it Phys. Lett.} {\bf B 389} (1996) 609.

\bibitem{lep6} P.~Cho, {\it Phys. Lett.} {\bf B 368} (1996) 171.

\bibitem{brch} E.~Braaten and Y.~Chen, {\it Phys. Rev. Lett.}
{\bf 76} (1996) 730.

\bibitem{trans} P.~Cho and M.~Wise, {\it Phys. Lett.}
{\bf B 346} (1995) 129;
S.~Baek, P.~Ko, J.~Lee and H.S.~Song, 
{\it Phys.Rev.} {\bf D 55} (1997) 6839; 
M.~Beneke and I.~Rothstein, {\it Phys. Lett.} {\bf B 372} (1996) 157; 
M.~Beneke and M.~Kr\" amer, {\it Phys. Rev.} {\bf D 55} (1997) 5269.

\bibitem{hadro} S.~Gupta and K.~Sridhar, {\it Phys. Rev.} {\bf D 54} 
(1996) 5545; 
M.~Beneke and I.~Rothstein, {\it Phys. Rev.} {\bf D 54} (1996) 2005; 
S.~Gupta and K.~Sridhar, {\it Phys. Rev.} {\bf D 55} (1997) 2650.

\bibitem{kim} C.S.~Kim, J.~Lee and H.S.~Song, 
KEK Preprint KEK-TH-474, hep-ph/9610294.

\bibitem{psigam} D.P.~Roy and K.~Sridhar, {\it Phys. Lett.} {\bf B 341} 
(1995) 413.

\bibitem{bfp} V.~Barger, S.~Fleming and R.J.N.~Phillips,
{\it Phys. Lett.} {\bf B 371} (1996) 111.

\bibitem{lhc} K.~Sridhar, {\it Mod. Phys. Lett.} {\bf A 11} 
(1996) 1555.

\bibitem{hera} R.M.~Godbole, D.P.~Roy and K.~Sridhar, 
{\it Phys. Lett.} {\bf B 373} (1996) 328.

\bibitem{cakr} 
N.~Cacciari and M.~Kr\"amer, 
{\it Phys. Rev. Lett.} {\bf 76} (1996) 4128;
P.~Ko, J.~Lee, H.S.~Song, 
{\it Phys. Rev.} {\bf D 54} (1996) 4312.

\bibitem{brw} M. Beneke, I.Z. Rothstein, M.B. Wise,  
{\it Phys. Lett.} {\bf B 408} (1997) 373.

\bibitem{onep} Y.-P.~Kuang, S.F.~Tuan and T.-M.~Yan,
{\it Phys. Rev.} {\bf D 37} (1988) 1210. 

\bibitem{e760} T.A.~Armstrong et al.,
{\it Phys. Rev. Lett.} {\bf 69} (1992) 2337. 

\bibitem{mine} K.~Sridhar, {\it Phys. Rev. Lett.} {\bf 77} (1996) 4880.

\end{thebibliography}
\end{document}